# Variability of Antenna Signals from Dust Impacts

Mitchell M. Shen[1,2,3], Zoltan Sternovsky[2,3], David M. Malaspina[2,4]

1. *Department of Astrophysical Sciences, Princeton University, Princeton, NJ 08544*
2. *Laboratory for Atmospheric and Space Physics, University of Colorado, Boulder, CO 80303*
3. *Smead Aerospace Engineering Sciences Department, University of Colorado, Boulder, CO 80303*
4. *Department of Astrophysical and Planetary Sciences, University of Colorado, Boulder, CO 80309*

**Key Points:**

(1) Antennas in dipole configuration are sensitive to dust impacts with the measured signals depending on impact location.

(2) Dust impacts at lower speeds produce complex and variable antenna signals, indicating the compound nature of impact-generated plasmas.

(3) Laboratory measurements performed with accelerated iron and aluminum particles generate similar antenna signals.

**Key Words:** cosmic dust, dust detection, antenna measurements

***Corresponding author:*** mitchellshen@princeton.edu

# Abstract

Electric field instruments carried by spacecraft are complementary to dedicated dust detectors by registering transient voltage perturbations caused by impact-generated plasma. The signal waveform contains information about the interaction between the impact-generated plasma cloud and the elements of spacecraft – antenna system. The variability of antenna signals from dust impacts has not yet been systematically characterized. A set of laboratory measurements are performed to characterize signal variations in response to spacecraft parameters (bias voltage and antenna configuration) and impactor parameters (impact speed and composition). The measurements demonstrate that dipole antenna configurations are sensitive to dust impacts and that the detected signals vary with impact location. When dust impacts occur at low speeds, the antennas typically register smaller amplitudes and less characteristic impact signal shapes. In this case, impact event identification may be more challenging due to lower signal-to-noise ratios and/or more variable waveforms shapes, indicating the compound nature of non-fully developed impact-generated plasmas. To investigate possible variations in the impacting materials, the measurements are carried out using two dust samples with different mass densities: iron and aluminum. No significant variations of the measured waveform or plasma parameters obtained from data analysis are observed between the two materials used.

# 1. Introduction

Electric field or plasma wave instruments using antennas can register the impacts of cosmic dust particles on spacecraft, as observed by a range of missions [*Babic et al.*, 2022; *Gurnett et al.,* 1983, 1987, 1997; *Kurth et al.,* 2006; *Meyer-Vernet et al.*, 2009, 2017; *Malaspina et al.*, 2014, 2020; *Ye et al.,* 2014, 2016a, 2016b, 2018, 2019, 2020; *Kellog et al.,* 2016; *Page et al.*, 2020; *Pusack et al.*, 2021; *Szalay et al.,* 2020; *Vaverka et al.,* 2018, 2019; *Zaslavsky et al.,* 2012, 2015, 2021]. A physical model based on first principles has been recently proposed to interpret and analyze dust impact waveforms recorded by antennas [*Shen et al.*, 2021a; 2021b]. In this model, the induced charging (and corresponding potential differences among spacecraft elements) from the expanding cloud of electrons and ions from the impact plasma are primarily responsible for the characteristic shapes of the impact signals. The model accounts for capacitive coupling between the spacecraft and the antenna elements and includes the discharge of the voltage signals through electric components and the plasma environment.

Impact plasma is the transient cloud of electrons and ions generated by the impact of a dust particle on a solid target surface (e.g., *Auer* [2001]). While the physical processes involved in the generation of impact plasmas are poorly understood, laboratory measurements revealed that the total generated charge approximately follows a power law, $Q_{IMP} \approx Q_i = |Q_e| = \gamma m v^\beta$, where $m$ is dust mass, and $v$ is the impact speed. Coefficients $\gamma$ and $\beta$ are characteristics of the target material and have been determined for various materials [e.g., *Auer,* 2001; *Collette et al.,* 2014; *Shen*, 2021c]. The impact plasma consists of free electrons, cations, and some fraction of anions. Other basic parameters of the impact plasma are the composition of ions and the energy distributions (or effective temperatures) of the charged species. The composition of ions depends on the dust and target materials and varies strongly with impact speed. Impact plasma ion composition has been studied in the laboratory using a range of dust materials and setups, where the ions are extracted from the impact plasma and subsequently examined using time-of-flight techniques [e.g., *Fiege et al.,* 2014; *Hillier et al.,* 2014; 2018, *Srama et al.*, 2009]. Generally, at high speeds (> 20 km/s), the ion composition is dominated by singly charged atomic species. On the other hand, higher-mass molecular and cluster ions are present in significant quantities at lower impact speeds. The effective temperatures of the electrons and ions are in the ranges of 1 – 4 eV and 4 – 15 eV, respectively, as determined from laboratory experiments for a small number of dust-target material combinations [*Collette et al.,* 2016; *Nouzák et al.*, 2020, *Kočiščák et al.*, 2020].

The effective temperatures are relevant for calculating the fraction of charge carriers collected by or escaping from a spacecraft.

The electrostatic model presented by *Shen et al.* [2021b] was in good agreement with experimental data collected in the laboratory using scaled-down spacecraft models. However, these laboratory measurements explored a limited parameter space of antenna configuration, impact speed range, and dust material. This study expands on these three specific parameters, as discussed below.

(1) Antenna configuration:

According to the electrostatic model, there are two dominant physical mechanisms for how impact plasma generates voltage signals on the spacecraft and antennas: charge recollection and induced charging. For the former, a fraction of the charge from the impact plasma is collected by spacecraft surfaces. For the latter, the escaping part of the impact plasma is responsible for generating the induced charging signal. The escape of electrons occurs over timescales that are often difficult to resolve with antenna electronics. However, the escape of ions is slower, and the corresponding induced charge is found to be primarily responsible for the characteristic shape and duration of impact signals. The duration scales inversely with the ion expansion speed (i.e., slower expansion results in longer duration). Observational evidence also suggests that ion escape occurs in the form of a diverging beam. When this beam passes over an antenna, that antenna observes an enhanced positive charge.

*Shen et al.* [2021b] demonstrated this effect for a monopole antenna configuration, i.e., measuring the potential difference between the antenna and the spacecraft. This study presents measurements with antennas configured as a dipole, i.e., measuring the potential difference between two antenna elements. Past studies suggested that antennas in a dipole configuration are much less sensitive to dust impacts compared to those operated as monopoles [e.g., *Tsintikidis et al.,* 1994; *Meyer-Vernet et al.,* 2009, 2014; *Ye et al.,* 2016, 2020, *Zaslavsky et al.,* 2021]. *Ye et al.* [2016] reported a result from the ring plane crossing of the Cassini spacecraft operating in the Saturnian system, where the antenna mode of operation was switched from monopole and dipole halfway through the crossing. The data clearly indicated significantly stronger dust impact signals in the monopole mode. The authors suggested that the dipole mode primarily detects impacts on the antenna booms rather than on the spacecraft. *Meyer-Vernet et al.* [2014] and *Zaslavsky et al.* [2021] reported the observation of dust impacts in dipole mode and their interpretation in terms of signals induced on the antenna by the electric field of the impact plasma. Laboratory simulation measurements were performed using a scaled-down model of the Cassini spacecraft by *Nouzák et al.* [2018]. This experimental campaign indicated that antennas operated in a dipole mode are insensitive to impacts on the spacecraft body or the monopole antenna. However, these measurements were limited to only a few impact locations relatively distant from the dipole antennas. On the other hand, a recent study by *Page et al.* [2020] reported that the antennas operated on the Parker Solar Probe mission were similarly sensitive to dust impacts both in dipole and monopole modes. In this study, we revise the experimental findings of *Nouzák et al.* [2018] and show that dipole antennas are sensitive to dust impacts if the impact location is such that the impact plasma expands over one of the antennas, and this expansion is asymmetric between the antenna pairs.

(2) Impact speed:

Observational evidence shows that there can be significant differences in the properties of the impact plasma (including the generated total charge) between individual dust impact events, even if the dust mass and impact velocity are similar. Such variations are less pronounced for speeds of 20 km/s, where the impact plasmas are relatively "well-behaved," meaning that the fitting of the measured antenna signals results in consistent fit parameters when the electrostatic model by *Shen et al.* [2021a, 2021b] is applied. This study reports a large variability in antenna signals at lower impact speeds, around 5 km/s. These measurements suggest that the large variability in the parameters of the impact plasmas may complicate the recognition of valid dust impact events and their statistical analysis.

The inherent problem of identifying dust impacts in antenna signal waveforms always complicates the data analysis. Theoretically, the characteristic signal shape (including a preshoot, main peak, and a discharging curve [*Nouzák* et al., 2018]) allows for the identification of whether an impact occurred or not. The first prerequisite to identification is that the impact signals must develop the described characteristic features and be above the signal-to-noise ratio threshold. Therefore, light impacts (e.g., small-momentum dust grains) or uncommon waveforms (saturated amplitude or wiggling ones) may be thrown away. Second, preshoot features may not always be registered due to "poorly-developed" impact plasma clouds or simply because some missions did not carry fast enough front-end electronics to capture fast electron escape. Third, variation in ambient plasma environments leads to diverse discharging time constants where the denser the plasma density, the more transient the signal will be and vice versa. A peak detector is commonly applied for dust impact identification. Proper signal filtering and empirical analysis would improve the discrimination in electric signals between transient dust impacts and plasma wave measurements. Intrinsically, extracting impact waveforms from background noise and distinguishing them from plasma wave measurements is the crux of the problem.

(3) Dust material:

Prior laboratory measurements were usually performed using an iron-tungsten dust-target material combination. This was done to eliminate a potential variable that could complicate the direct comparison of results between experimental campaigns. A unique set of laboratory measurements were performed using an aluminum-tungsten material combination to show that the results and electrostatic model of *Shen et al.* [2021a; 2021b] are not limited to a single unique material combination.

## 2. Experimental Setup

The experimental setup used for the measurements described below is like that used in previous investigations [*Shen et al.*, 2021b]. Briefly, the spacecraft (SC) is modeled in the laboratory as a conductive sphere that is $R_{SC} = 7.5$. cm in diameter (Fig. 1). Four antennas are mounted on the model SC in one plane and spaced 90° apart. The antennas are $L_{ANT} = 27$ cm long and 1.6 mm in diameter. The spherical body and four antennas are made of stainless steel, while the former is coated with graphite paint to provide uniform potential on its surface [*Robertson et al.*, 2004]. Organic solvents were used for cleaning the non-coated surfaces. A strip of tungsten (W) foil is wrapped around the circumference of the model SC in the same plane as the antennas are mounted. The W is used as the target for the dust impacts to make the measurements directly comparable to prior results reported by *Nouzák et al.* [2018, 2020] and *Shen et al.* [2021a, 2021b]. The model SC is installed onto a vertical rotary shaft in the center of a large vacuum chamber (1.2 m in diameter, 1.5 m in length), and the experiments are performed in high-vacuum, ~$10^{-6}$ Torr, without a background plasma.

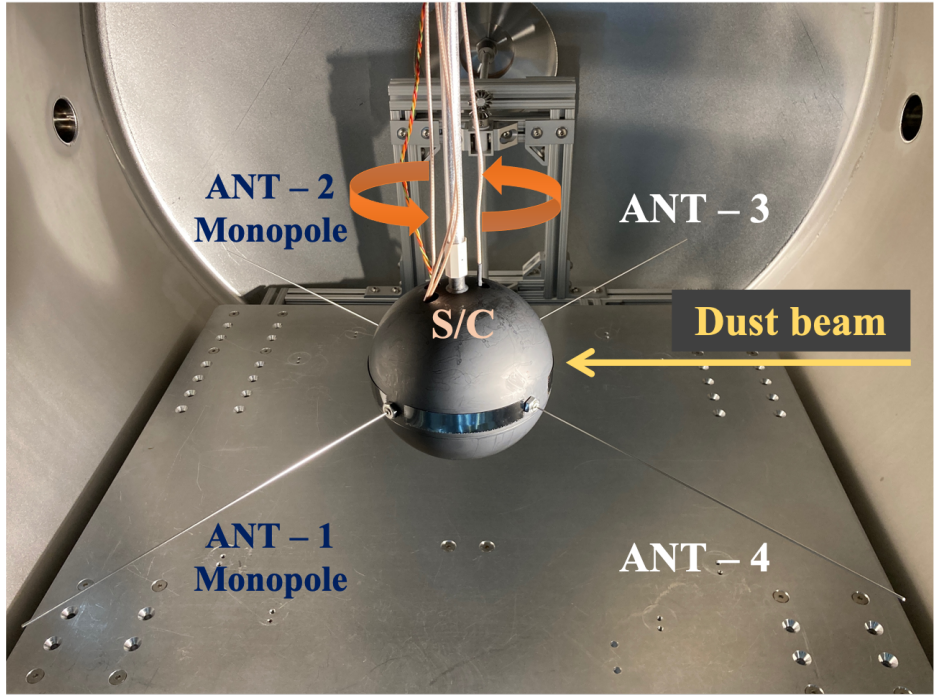

**Figure 1:** Spherical spacecraft model with the four cylindrical antennas mounted inside the vacuum chamber. See text for detail.

The four antennas are configured as one dipole pair ($V_{meas} = V_{ANT-3} - V_{ANT-4}$) and two independent monopoles ($V_{meas} = V_{ANT-1} - V_{SC}$ or $V_{ANT-2} - V_{SC}$). In the former case, the signal measured is the voltage difference between the antennas, while in the latter case, the measured signal is with respect to the SC body. The electronics are three channels of instrumentation amplifiers housed within the SC's spherical body. The amplifiers operate with a voltage gain of 50, have a bandwidth of 270 Hz – 5 MHz, and are described in more detail in *Shen et al.* [2021a]. Each element of the lab model, i.e., the SC body and the four antennas, can be biased independently through large-value resistors, $R_{bias,SC} = 2.5\ M\Omega$ and $R_{bias,ANT} = 5M\Omega$, respectively [*Shen et al.,* 2021b]. In this study, the SC and antennas are biased at the same potential. The bias resistors provide a discharge path for each of the elements. When combined with the respective effective capacitances of the SC ($C_{eff,SC} \approx 42$ pF) and antennas ($C_{eff,ANT} \approx 16$ pF), the characteristic RC discharge time constants can be calculated as $R_{bias,SC}C_{eff,SC} = 105$ μs and $R_{bias,ANT}C_{eff,ANT} = 80$ μs. The calculation of effective capacitances is provided in *Shen et al.* [2021b]. The waveforms measured by the three amplifiers are recorded using a fast-digitizing oscilloscope.

The dust accelerator facility operated at the University of Colorado is used to provide submicron-sized particles in a velocity range of about 1 – 40 km/s [*Shu et al.*, 2012]. Both iron (Fe) and aluminum (Al) dust samples were used. Measurements with the Fe-W (dust-target) material combination provide a direct comparison with the studies conducted by *Collette et al.* [2015; 2016], *Nouzák et al.* [2018; 2020], and *Shen et al.* [2021a; 2021b]. Measurements with Al dust were added to this study to investigate variation with impactor materials.

## 3. Experimental Data

### 3.1 Signal Variation with Antenna Configuration

Figure 2 shows an overview of typical antenna signals measured for different impact locations between antennas #3 and #4. The signals shown are for two monopole antennas (#1 and #2) and one dipole pair, where the measured signal is $V_{meas} = V_{ANT-3} - V_{ANT-4}$ (see Fig. 1). The columns correspond impact locations 10°, 30°, and 45° measured from antenna #3, while the rows are for different applied bias potentials, 0 V, +5 V, and – 5 V. The same bias potential is applied to all elements, i.e., the spacecraft body and the four antennas. The limitations in terms of the bias potential investigated are due to the availability of the dust accelerator. The +5 V bias potential represents the typical case of a SC in interplanetary space. All collected impact events are for Fe-W dust-target material combinations and impact velocities ≥ 20 km/s.

The monopole signals for 0 V and +5 V bias voltages are characteristic shapes described in detail by the electrostatic model by *Shen et al.* [2021b]. Briefly, the signal starts with a sharp negative spike due to the fast escape of free electrons that leave the spacecraft with a net positive charge. The following slower positive rise is from the escape of the slower ions that charge the spacecraft negatively. The time constant of this rise is on the order of $R_{SC}/v_i$, where $v_i \approx 10$ km/s is the typical ion expansion speed [e.g., *Shen et al.,* 2021a]. The subsequent slow decay is driven by discharging of the spacecraft through the biasing resistors of the electronics (Sec. 2).

The common observation at high impact speeds is that the total charge of escaping electrons at 0 V bias is about half of those escaping ions. This results in a non-zero total collected charge and the characteristic shape of the antenna signals often observed by instruments in space. The physical explanation for this fact is that the free electrons acquire an isotropic distribution during the early phases of impact plasma expansion, and thus half of the electrons are naturally recollected by the spacecraft as the plasma cloud expands [e.g., *Shen et al.*, 2021a, 2021b]. The main difference between the 0 V and +5 V bias voltage cases is that the relative amplitude of the sharp negative spike is smaller for the +5 V bias. Since the typical effective ion temperatures are larger than 5 eV [*Collette et al.,* 2016; *Nouzák et al.*, 2020, *Kočiščák et al.*, 2020], the ion-dominated part of the waveform for the +5 V bias potential is similar to that of the case with 0 V bias. The physical explanation is that a smaller fraction of electrons can escape a spacecraft with a positive bias potential. For the –5 V bias potential, on the other hand, a larger fraction of electrons can escape, while some fraction of the ions is recollected by the spacecraft [see also *Nouzák et al.*, 2018; 2020].

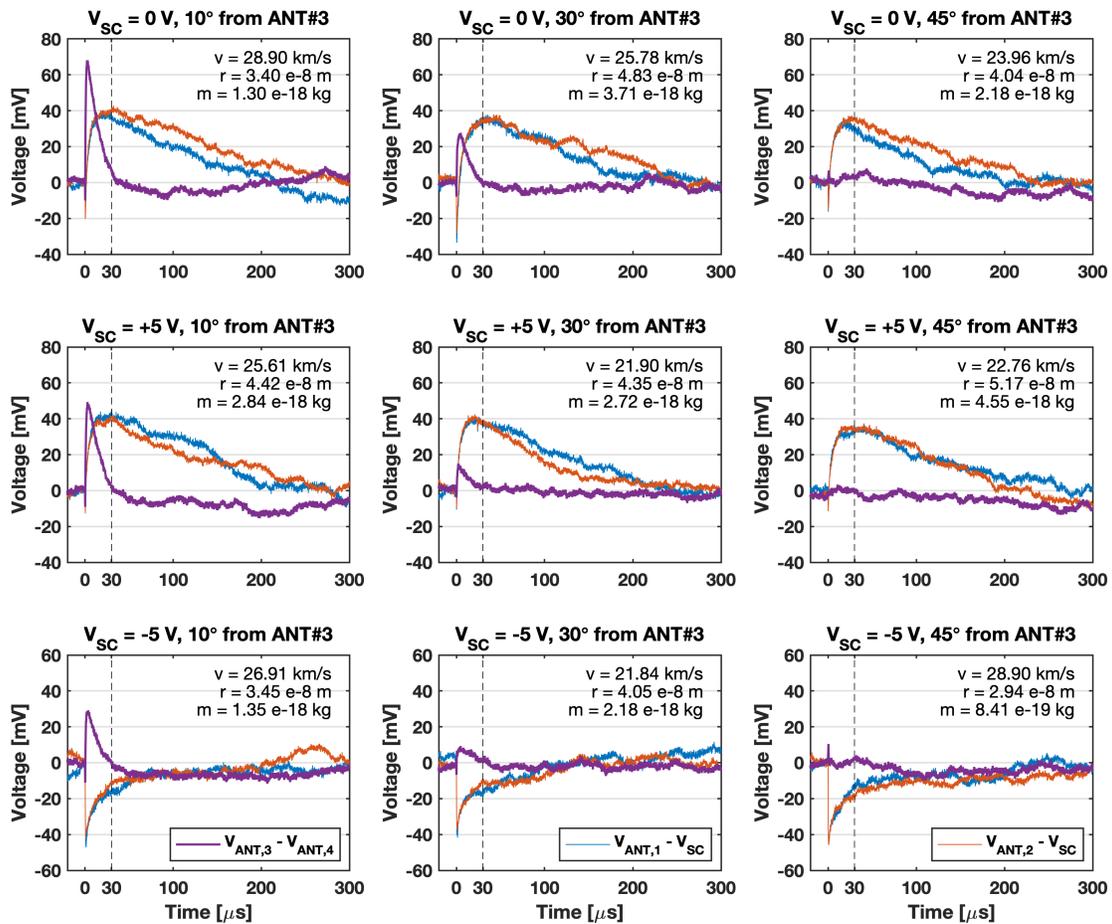

**Fig. 2:** Dipole (purple) and monopole (blue and orange) antenna signals measured in the laboratory for three impact locations (10°, 30°, and 45° from left to right column) and three bias voltages (0V, +5V, and -5V from top to bottom row). The impact speed (*v*), dust radius (*r*), and mass (*m*) of iron dust particles are provided for individual panels. The vertical dashed lines mark 30 μs after the impact.

The signals from the two monopole antennas are very similar to one another. This is because the impact locations are far from either antenna, and the expanding impact plasmas have little interaction with these antenna elements. In other words, the signals on these monopole antennas are dominated by spacecraft charging. The dipole signals are most pronounced for the 10° impact location, where the expansion of the positive ion cloud over antenna #3 is responsible for the positive amplitude of the signal. In this most pronounced case (10° impact location, 0 V or +5 V bias potentials), the amplitude of the dipole signal can be just as large or even larger than the monopole signals. This is because the effective capacitance of the antenna elements is lower than that of the spacecraft; thus, even a small, induced charge on the antennas can generate significant voltage signals.

Three further interesting observations can be made from the measured dipole signals. The first is their characteristic duration of about 30 μs for all cases shown in Fig. 2. The time constant is determined by the length of the antenna and the ion expansion speed, $L_{ANT}/v_i$. The second one is that the dipole signal is diminishing as the relative distance of the impact location from antenna #3 is increasing. At the 45° impact location, i.e., exactly in between the dipole antennas, the signal is consistently close to zero. This fact implies that the expanding cloud of ions from the impact plasma is approximately symmetrical, at least for the normal impact directions investigated in this study. It is noted that the surface roughness of the W target strips, and the size of the dust particles are of a similar order of magnitude (~ 50 nm). The third observation is that the sharp negative spike is still observable for the 10° impact location but mostly disappears at 45°. The residual sharp peaks in the latter case could be caused by asymmetries in the electron cloud expansion or possibly slight differences in the antenna geometries and their mounting. For monopole antennas, this feature is due to the fast-escaping electrons, as discussed above. However, for dipole antennas, one would expect such a feature to disappear due to the symmetry between the antennas. A feasible explanation is that for the 10° dipole case, the antenna senses the induced charge directly from the escaping electrons rather than the corresponding charge sensed by the SC.

## 3.2    Characteristics of Low-Speed Dust Impacts

Figure 3 shows a set of monopole signals measured upon Fe dust impacting at a lower impact speed, around 5 km/s. The top three rows demonstrate different applied bias potentials of 0 V, +5

V, and –5 V; two individual events are attached for each biasing. The impact location is 45° between the two monopole antennas #1 and #2. Due to the geometric symmetry of the model SC (see Fig.1), ideally, the two monopole signals should be qualitatively similar.

There are similarities and differences between high-speed and low-speed impact signals. In general signal characteristics, the fast negative-going signal indicates the escape of free electrons, followed by the slower escape of ions and restoration by discharging. The first difference is that at low impact speeds, the amounts of escaping electrons and ions are approximately equal, as indicated in the case of 0 V bias. The possible explanations are that the electrons of the impact plasma are no longer isotropic in their velocity distribution or not all ions are escaping. As the net collected charge is close to zero, the duration of the impact signal is thus determined by the dynamics of the ions ($\sim R_{SC}/v_i$), not the characteristic discharge of the spacecraft.

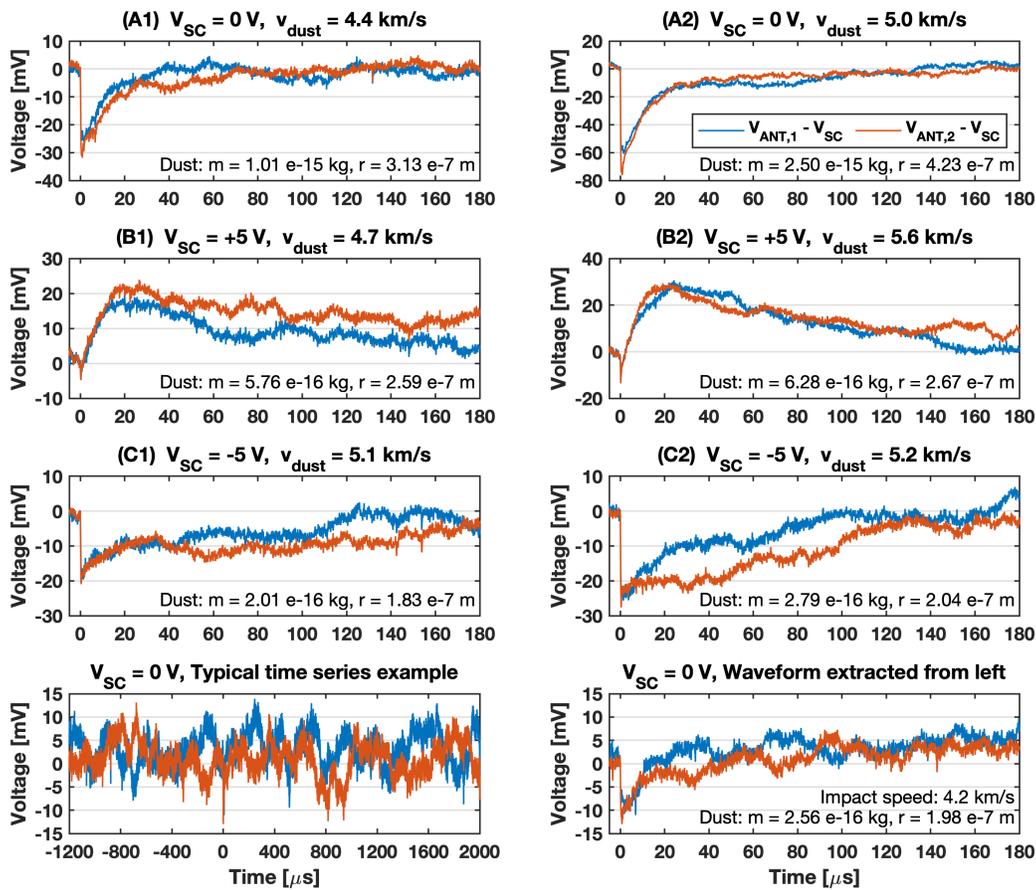

**Fig. 3:** Typical monopole waveforms measured in the laboratory for low-speed particles ($\cong$ 5 km/s) where the impact location is at 45° between antenna #1 and #2 for three bias voltages. The mass, size, and velocity of the dust particles are provided for individual events.

The lack of a distinct main peak in the signal leads to difficulties identifying impact occurrence. On the other hand, applying bias voltages changes the balance between the ratio of escaping electron/ion charges resulting in a nonzero net collected charge. Note that applying +5 V bias voltage results in the recollection of a larger fraction of the electrons by the SC and thus shifting the waveform upward, with less pronounced electron escape signals and more dominant signals due to the escaping ions. This makes the signals more similar to the case with high-speed dust impacts. The observed complex and variable shapes are driven by the compound nature of impact-generated plasmas, yet the role of anions has not been systematically discussed. Nonetheless, the production of anions is considered a minor effect [*Kočiščák et al.*, 2020 and references therein].

Example waveforms from an impact event that generated less impact charge and had a low signal-to-noise ratio are presented in the bottom panels. The bottom right panel shows a subset of the impact waveform from the bottom left panel. The onset time of dust impact is known in laboratory experiments but would be unknown in space measurements. This demonstrates the difficulties of dust impact identification without characteristic signal features. For instance, the leading edge (pre-spikes) can be considered as a trigger in signal identification; however, it strongly relies on the electronics' response.

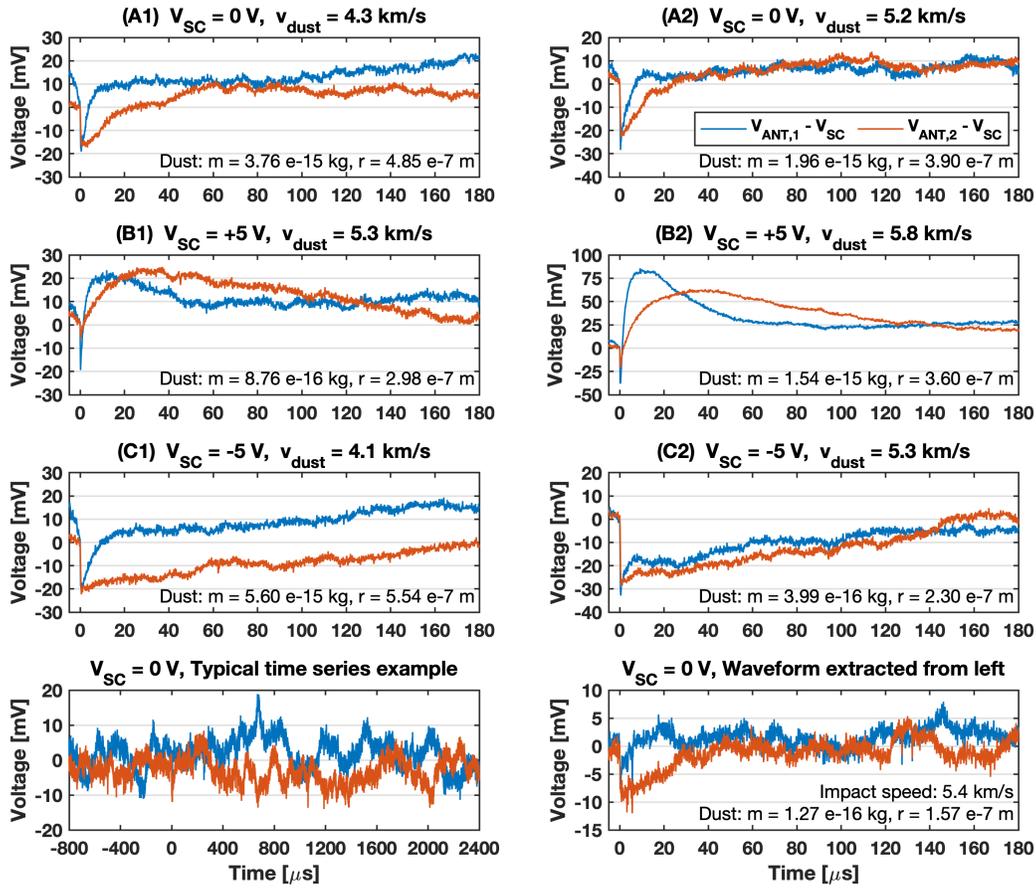

**Fig. 4:** Typical monopole waveforms measured in the laboratory for 10° from antenna #1 for three different bias voltages. The properties of slow dust particles ($\cong$ 5 km/s) are labeled.

Figure 4 shows the typical monopole signals measured on antennas #1 and #2 when the impact location is 10° from antenna #1. There are large variations between individual events. First, the amplitudes of negative-going preshoots are different in the two monopole signals due to the extra induced charging of escaping electrons on nearby antenna #1, as discussed in *Shen et al.* [2021b]. Second, in the top six panels, the main peak signal rise for each event (induced charging from escaping ions) on monopole #1 appeared earlier than that on monopole #2, with a time difference less than 30 μs. Note that this 30 μs represents a characteristic timescale for the expanding plasma cloud that passes through the antenna by considering a typical ion expansion speed of approximately 10 km/s. However, it is unlikely that ions generated by low-speed impacts may move greater than 10 km/s. Instead, it implies that the angular distribution of the ion plume may be more divergent in space than those generated by higher velocity impacts (≥ 20 km/s). With this interplay between the angular expansion of impact plasma cloud and the geometry of antennas, such an early enhancement of the main peak coupled with the described equal amount phenomena of escaping charged particles shortens its characteristic time constant and deviates from the characteristic signal shape, thus making impact identification and the more detailed analysis of the waveforms challenging [e.g., *Shen et al.* 2021a; 2021b].

Generally, dust impacts with low speed produce smaller signal amplitude due to less impact charge generated and non-fully developed impact plasma upon impact, thus introducing difficulties in extracting them from the background noise. It is noteworthy that the impact occurrences can still be identified as wideband noise through the power spectrum in antenna signals [*Aubier et al.*, 1983; *Meyer-Vernet et al.*, 2009]. However, this form of the signal provides limited information on impactors.

## 3.3    Signal Variation with Dust Composition

A unique set of measurements were performed using Al-W dust-target material combinations, and impact velocities are limited to ≥ 20 km/s in order to compare with prior studies of Fe dust impacts performed by *Nouzák et al.* [2018, 2020] and *Shen et al.* [2021a, 2021b]. The presented monopole signals in Figure 5 span two impact locations (45° and 10° apart from antenna #1) and three bias voltages applied on the SC (0V, +5V, and -5V).

Signal features on individual events, including preshoot and the main peak of ion cloud expansion followed by the SC discharging, appear qualitatively similar to the Fe dust impact measurements reported in *Shen et al.* [2021b] and references therein. The signal variation regarding the bias voltages and impact locations where the induced charging on antenna led by ion cloud expansion (see main peaks in monopole #1, especially 10° column) are also remarkably alike. The comprehensive characterization of signal generation mechanisms has been discussed in *Shen et al.* [2021a, 2021b].

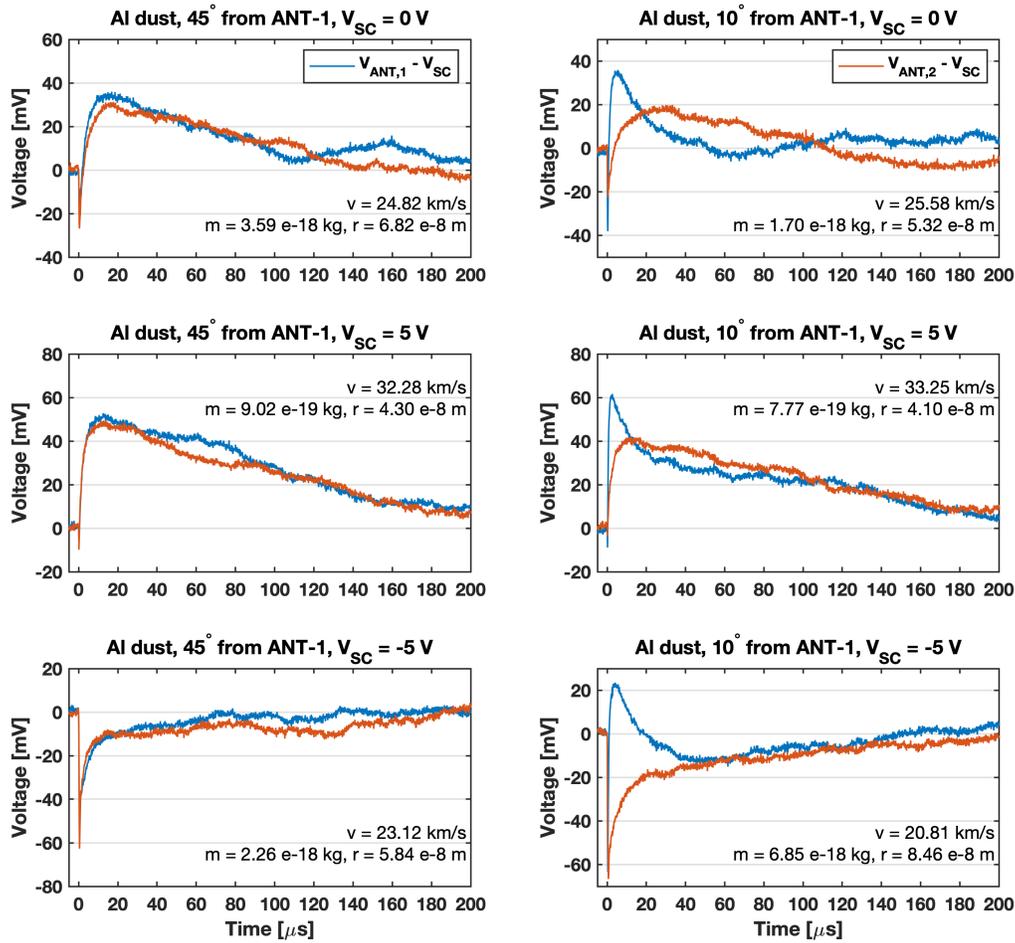

**Fig. 5:** Typical monopole waveforms measured in the laboratory for 45° (left column) and 10° (right column) from antenna #1 under three different bias voltages. The dust-target combination is Aluminum to Tungsten (Al-W). The properties of dust particles are denoted.

A model fitting example of Al dust impacting 45° between antennas #1 and #2 is provided in Figure 6 (extracted waveform from the top left panel in Fig. 5). The bias voltage on SC and four antennas is set to 0V. The two monopole signals are not identical even though the impact location is directly 45° in between. It is a superposition effect of (1) different antenna capacitances of $C_{ANT,1} = 9.5$ pF and $C_{ANT,2} = 10.5$ pF, and (2) the geometry of the impact plasma cloud expansion. Using the electrostatic model presented in *Shen et al.* [2021b], the capacitive coupling of elements has been considered in model fitting through industry-standard SPICE (Simulation Program with Integrated Circuit Emphasis) software using an electronics circuit diagram. The resulting fitting parameters are impact charge ($Q_{IMP}$), ion expansion speed ($v_i$), geometric coefficient ($\kappa$), and auxiliary parameters of negative (subscript $e$) and positive (subscript $i$) amplitudes on monopole amplitudes ($\zeta_{e,ANT-1}, \zeta_{i,ANT-1}, \zeta_{e,ANT-2}, \zeta_{i,ANT-2}$) [*Shen et al.,*2021b]. These auxiliary parameters serve to compensate for the induced charging on a specific antenna regarding the angular distribution of impact plasma cloud expansion.

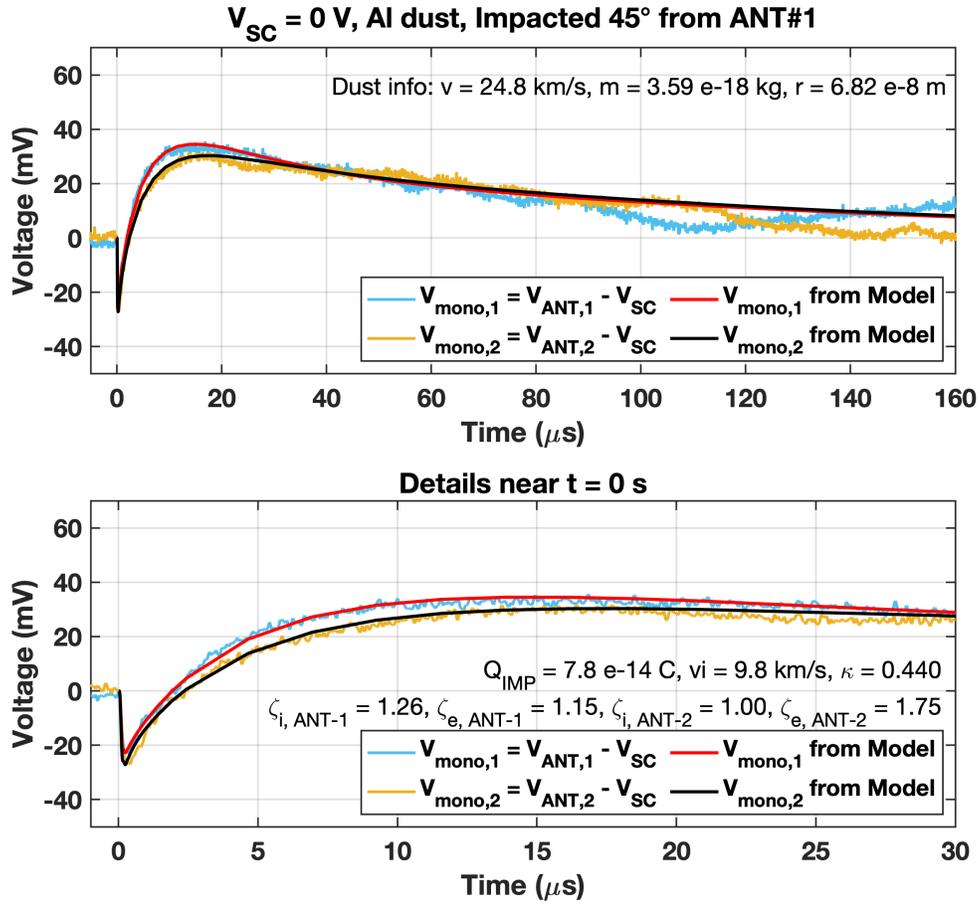

**Fig. 6:** Monopole waveforms measured in the laboratory with Al dust impacting the 45° location from antenna #1 (same top left panel in Fig. 5). The dust particle properties are provided in the top panel, while the detailed early phases of the impact plasma expansion and fitting parameters are labeled in the bottom one.

The obtained fitting parameters of the impact plasma are $Q_{IMP} = 7.8 \times 10^{-14}$ C and $v_i = 9.8$ km/s under two control variables $\kappa = 0.44$ and $v_e = 10^3$ km/s. The $\kappa$ coefficient represents the field of view from the impact site where the isotropic electrons can escape unobscured. It is considered a SC surface property and set to be the same values provided in *Shen et al.* [2021b] (same model SC and impact location, c.f., Figure 6 in that article). The escape speed of electrons $v_e$ is set to a proper thermal speed of 2 eV [*Shen et al., 2021a*]. These two fitting values, $Q_{IMP}$ and $v_i$, fall within the reasonable range that agree with prior studies.

Auxiliary parameters are introduced to optimize the fits. Each represents the expansion behaviors of escaping electrons and cations registered from the aspect of a specific antenna. Results show that $\zeta_{e,ANT-1} = 1.15$ and $\zeta_{i,ANT-1} = 1.26$ for monopole #1 while $\zeta_{e,ANT-2} = 1.75$ and $\zeta_{i,ANT-2} = 1$ for monopole #2. These values fall within the reasonable ranges of $\zeta \geq 1$ at 45° impact location compared to those presented in *Shen et al.* [2021b]. Note that the existing model simplifies the escaping charged particles as point sources moving radially away from the impact site with constant velocities. However, the impact-generated electrons and cations would expand in isotropic and conical distributions, respectively. Hence, these auxiliary parameters are fitted $\geq 1$ for reconciling the underestimation of induced charging on the antennas regarding solid angles in the electrostatic model.

The impact signals and model fitting indicate no significant variations regarding the dust composition. This result is consistent with studies by *Auer and Sitte* [1968] and *Adams and Smith* [1971], who found that the impact response is mainly associated with the target material rather than the composition of impactors.

# 4. Summary and Conclusions

The article aims to characterize experimentally the variability of antenna signals created by plasma, generated by dust impacts, based on impact location, SC potential, antenna configurations, dust composition, and impact speeds. A spherical model SC has been used for laboratory measurements. Test conditions include (a) monopole and dipole configurations response to impact locations and SC potential with hypervelocity dust (≥ 20 km/s), (b) signal features under low-speed dust impacts (∼ 5 km/s), and (c) antenna signal characteristics with a different dust composition (Al) at hypervelocity. Features of signal waveforms are qualitatively characterized, and a demonstration of model fitting is performed.

Recent laboratory [*Nouzák et al.*, 2018] and data analysis studies [*Ye et al.*, 2016b; 2020] suggested that antennas in dipole configuration might be insensitive to dust impact plasmas. The present work revises conclusions from the former and demonstrates experimentally that antennas in a dipole configuration indeed are sensitive to impacts, provided that the impact plasma cloud passes close to one antenna in the pair. Electrostatic induction of charges from the impact plasma was proposed as the mechanism that generates the voltage signals on the antenna by *Meyer-Vernet et al.* [2014]. The detailed model of such interactions was provided by *Shen et al.* [2021b]. A cross-comparison between dipole and monopole dust detection in the laboratory shows that the former is useful for the determination of impact location and source categorization, while the latter provides detailed impact plasma parameters for mass and size determination. It is suggested that both types of measurements should be taken together for comprehensive dust studies using electric field or plasma wave instruments.

Small amplitude and more complex impact signals were observed under lower-speed dust impacts (∼5 km/s) due to non-fully developed impact plasma clouds. The resultant waveforms may be challenging to discriminate from the background noise. Signals with equal amounts of escaping electrons and ions when SC at 0 V bias suggest the electrons generated upon impact are no longer distributed isotropically in velocity or not all ions are energetic enough to escape. It is speculated that the voltage developed on the SC might alter the signal features more significantly than that in high velocity impact events (≥ 20 km/s). A detailed waveform analysis may improve dust detection efficiency but still lacks a comprehensive characterization of impact plasma generated by low-speed impacts.

Aside from the typical Fe dust particles, Al was chosen to verify whether the impact waveforms would vary with the composition of the dust material (and thus the composition of the dust impact plasma) or dust density. The similarities of the data using iron and aluminum dust particles, and the corresponding data analysis concludes that the results are not unique to the iron-tungsten combination of dust-target materials that have been used in several of the prior experimental studies. This finding is consistent with prior work by *Auer and Sitte* [1968] and *Adams and Smith* [1971].

# Acknowledgment

Authors M.S. and Z.S.'s contributions to this study were supported by NASA's Cassini Data Analysis Program (CDAP), Grant NNX17AF99G. The contribution from author D.M. was supported by NASA contract NNN06AA01C. This publication is also supported by NASA's Heliophysics Guest Investigators – Open program (HGIO), Grant 80NSSC22K0753, which funds authors Z.S. and D.M. The operation of the dust accelerator was supported by NASA's Solar System Exploration Research Virtual Institute (SSERVI) Cooperative Agreement Notice, Grant 80NSSC19M0217. The authors thank John Fontanese for operating the dust accelerator during the experimental campaigns. The authors appreciate Craig Joy from the physics department supporting the machining of the spherical spacecraft model.

# Data Availability Statement

The laboratory data (Shen et al., 2022, Supporting Information data set) are publicly available in Zenodo repository (https://doi.org/10.5281/zenodo.7047232).

# Author contributions

Conceptualization – Z. Sternovsky

Formal analysis – M. Shen

Funding acquisition - Z. Sternovsky, D. Malaspina

Investigation – M. Shen

Methodology – M. Shen, Z. Sternovsky

Validation – M. Shen, Z. Sternovsky, D. Malaspina

Writing – original draft – M. Shen, Z. Sternovsky

Writing – review & editing – M. Shen, Z. Sternovsky, D. Malaspina

Manuscript accepted online by JGR: Space Physics on 22 March 2023.    https://doi.org/10.1029/2022JA030981# References

Adams, N. G., & Smith, D. (1971). Studies of microparticle impact phenomena leading to the development of a highly sensitive micrometeoroid detector. Planetary and Space Science, 19(2), 195-204.

Aubier, M.G., Meyer-Vernet, N. and Pedersen, B.M. (1983), Shot noise from grain and particle impacts in Saturn's ring plane. Geophys. Res. Lett., 10: 5-8. https://doi.org/10.1029/GL010i001p00005

Auer, A., & Sitte, K. (1968). Detection technique for micrometeoroids using impact ionization. Earth and Planetary Science Letters, 4(2), 178-183.

Auer, S., (2001). Instrumentation, in Interplanetary Dust, edited by E. Grün, B.A.S. Gustafson, S. Dermott, and H. Fechtig, pp. 385–444, Springer, New York.

Babic, K. R., Zaslavsky, A., Issautier, K., Meyer-Vernet, N., & Onic, D. (2022). An analytical model for dust impact voltage signals and its application to STEREO/WAVES data. Astronomy & Astrophysics, 659, A15.

Collette, A., Grün, E., Malaspina, D., and Sternovsky, Z. (2014), Micrometeoroid impact charge yield for common spacecraft materials, J. Geophys. Res. Space Physics, 119, 6019–6026, doi:10.1002/2014JA020042.

Collette, A., Meyer, G., Malaspina, D., and Sternovsky, Z., 2015. Laboratory investigation of antenna signals from dust impacts on spacecraft. J. Geophys. Res.: Space Physics 120 (7), 5298–5305.

Collette, A., Malaspina, D., and Sternovsky, Z., (2016). Characteristic temperatures of hypervelocity dust impact plasmas. J. Geophys. Res.: Space Physics 121 (9), 8182–8187.

Fiege, K., Trieloff, M., Hillier, J. K., Guglielmino, M., Postberg, F., Srama, R., ... & Blum, J. (2014). Calibration of relative sensitivity factors for impact ionization detectors with high-velocity silicate microparticles. Icarus, 241, 336-345.

Gurnett, D. A., Grün, E., Gallagher, D., Kurth, W. S., & Scarf, F. L. (1983). Micron-sized particles detected near Saturn by the Voyager plasma wave instrument. Icarus, 53(2), 236–254. https://doi.org/10.1016/0019-1035(83)90145-8

Gurnett, D. A., Kurth, W. S., Scarf, F. L., Burns, J. A., Cuzzi, J. N., & Grün, E. (1987). Micron-sized particle impacts detected near Uranus by the Voyager 2 plasma wave instrument. Journal of Geophysical Research: Space Physics, 92(A13), 14959-14968.

Gurnett, D. A., Ansher, J. A., Kurth, W. S., & Granroth, L. J. (1997). Micron-sized dust particles detected in the outer solar system by the Voyager 1 and 2 plasma wave instruments. Geophysical Research Letters, 24(24), 3125–3128. https://doi.org/10.1029/97gl03228

Hillier, J. K., Sternovsky, Z., Armes, S. P., Fielding, L. A., Postberg, F., Bugiel, S., ... & Trieloff, M. (2014). Impact ionisation mass spectrometry of polypyrrole-coated pyrrhotite microparticles. Planetary and Space Science, 97, 9-22.

Hillier, J. K., Sternovsky, Z., Kempf, S., Trieloff, M., Guglielmino, M., Postberg, F., & Price, M. C. (2018). Impact ionisation mass spectrometry of platinum-coated olivine and magnesite-dominated cosmic dust analogues. Planetary and Space Science, 156, 96-110.

Kellogg, P.J., Goetz, K., Monson, S.J., 2016. Dust impact signals on the Wind spacecraft. J. Geophys. Res. 121 (2), 966–991. https://doi.org/10.1002/2015JA021124.
21

Kočiščák, S., Fredriksen, Å., DeLuca, M., Pavlů, J., & Sternovsky, Z. (2020). Effective temperatures of olivine dust impact plasmas. IEEE Transactions on Plasma Science, 48(12), 4298-4304.

Kurth, W.S., Averkamp, T.F., Gurnett, D.A., Wang, Z., 2006. Cassini RPWS observations of dust in Saturn's E ring. Planet. Space Sci. 54 (9–10), 988–998.

Malaspina, D. M., Horányi, M., Zaslavsky, A., Goetz, K., Wilson, L. B., & Kersten, K. (2014). Interplanetary and interstellar dust observed by the Wind/WAVES electric field instrument. Geophysical Research Letters, 41(2), 266–272. https://doi.org/10.1002/2013gl058786

Malaspina, D. M., Szalay, J. R., Pokorný, P., Page, B., Bale, S. D., Bonnell, J. W., ... & MacDowall, R. J. (2020). In Situ Observations of Interplanetary Dust Variability in the Inner Heliosphere. The Astrophysical Journal, 892(2), 115.

Meyer-Vernet, N., Lecacheux, A., Kaiser, M. L., and Gurnett, D. A. (2009), Detecting nanoparticles at radio frequencies: Jovian dust stream impacts on Cassini/RPWS, Geophys. Res. Lett., 36, L03103, doi:10.1029/2008GL036752.

Meyer-Vernet, N., Moncuquet, M., Issautier, K., & Lecacheux, A. (2014). The importance of monopole antennas for dust observations: Why Wind/WAVES does not detect nanodust. Geophysical Research Letters, 41(8), 2716-2720.

Meyer-Vernet, N., Moncuquet, M., Issautier, K., and Schippers, P. (2017), Frequency range of dust detection in space with radio and plasma wave receivers: Theory and application to interplanetary nanodust impacts on Cassini, J. Geophys. Res. Space Physics, 122, 8–22, doi:10.1002/2016JA023081.

Nouzák, L., Hsu, S., Malaspina, D., Thayer, F. M., Ye, S.–Y., Pavlů, J., Němeček, Z., Šafránková, J. and Sternovsky, Z., 2018. Laboratory modeling of dust impact detection by the Cassini spacecraft. Planet. Space Sci.156, 85–91.

Nouzák, L., Sternovsky, Z., Horányi, M., Hsu, S., Pavlů, J., Shen, M. H., & Ye, S. Y. (2020). Magnetic field effect on antenna signals induced by dust particle impacts. Journal of Geophysical Research: Space Physics, 125(1), e2019JA027245.

Page, B., Bale, S. D., Bonnell, J. W., Goetz, K., Goodrich, K., Harvey, P. R., ... & Szalay, J. R. (2020). Examining dust directionality with the parker solar probe FIELDS instrument. The Astrophysical Journal Supplement Series, 246(2), 51.

Pusack, A., Malaspina, D. M., Szalay, J. R., Bale, S. D., Goetz, K., MacDowall, R. J., & Pulupa, M. (2021). Dust Directionality and an Anomalous Interplanetary Dust Population Detected by the Parker Solar Probe. The Planetary Science Journal, 2(5), 186.

Robertson, S., Sternovsky, Z., & Walch, B. (2004). Reduction of asymmetry transport in the annular Penning trap. *Physics of Plasmas*, *11*(5), 1753–1756. http://doi.org/10.1063/1.1688337

Shen, M. M., Sternovsky, Z., Horányi, M., Hsu, H.-W., & Malaspina, D. M. (2021a). Laboratory study of antenna signals generated by dust impacts on spacecraft. Journal of Geophysical Research: Space Physics, 126, e2020JA028965. https://doi.org/10.1029/2020JA028965

Shen, M. M., Sternovsky, Z., Garzelli, A., & Malaspina, D. M. (2021b). Electrostatic model for antenna signal generation from dust impacts. Journal of Geophysical Research: Space Physics, 126, e2021JA029645. https://doi.org/10.1029/2021JA029645

Shen, M. (2021c). Cosmic dust detection by antenna instruments - modeling and laboratory measurements (Order No. 28651348). Available from ProQuest Dissertations & Theses

Global. (2572580419). Retrieved from https://www.proquest.com/dissertations-theses/cosmic-dust-detection-antenna-instruments/docview/2572580419/se-2

Shen, Mitchell M., Sternovsky, Zoltan, & Malaspina, David M. (2022). Supplementary data to "Variability of Antenna Signals from Dust Impacts" [Data set]. Zenodo. https://doi.org/10.5281/zenodo.7047232

Shu, A., Collette, A., Drake, K., Grun, E., Horányi, M., Kempf, S., Mocker, A., Munsat, T., Northway, P., Srama, R., Sternovsky, Z., and Thomas, E., 2012. 3 MV hypervelocity dust accelerator at the Colorado Center for Lunar Dust and Atmospheric Studies. Review of Scientific Instruments 83 (7), 075108.

Srama, R. *et al.* Mass spectrometry of hyper-velocity impacts of organic micrograins. *Rapid Communications in Mass Spectrometry* **23**, 3895–3906 (2009).

Szalay, J. R., Pokorný, P., Bale, S. D., Christian, E. R., Goetz, K., Goodrich, K., ... & McComas, D. J. (2020). The Near-Sun Dust Environment: Initial Observations from Parker Solar Probe. The Astrophysical Journal Supplement Series, 246(2), 27.

Tsintikidis, D., Gurnett, D., Granroth, L. J., Allendorf, S. C., & Kurth, W. S. (1994). A revised analysis of micron-sized particles detected near Saturn by the Voyager 2 plasma wave instrument. Journal of Geophysical Research: Space Physics, 99(A2), 2261-2270.

Vaverka, J., Nakamura, T., Kero, J., Mann, I., Spiegeleer, A. D., Hamrin, M., Norberg, C., Lindqvist, P., & Pellinen-Wannberg, A. (2018). Comparison of Dust Impact and Solitary Wave Signatures Detected by Multiple Electric Field Antennas Onboard the MMS Spacecraft. Journal of Geophysical Research: Space Physics, 123(8), 6119–6129. https://doi.org/10.1029/2018ja025380

Vaverka, J., Pavlů, J., Nouzák, L., Šafránková, J., Němeček, Z., Mann, I., Ye, S., & Lindqvist, P. (2019). One-year analysis of dust impact-like events onto the MMS spacecraft. Journal of Geophysical Research: Space Physics. https://doi.org/10.1029/2019ja027035

Ye, S.-Y., Gurnett, D. A., Kurth, W. S., Averkamp, T. F., Kempf, S., Hsu, H. W., Srama, R. and Gruen, E., 2014. Properties of dust particles near Saturn inferred from voltage pulses induced by dust impacts on Cassini spacecraft. J. Geophys. Res.: Space Physics 119 (8), 6294–6312.

Ye, S.-Y., Gurnett, D. A., Kurth, W. S., 2016a. In-situ measurements of Saturn's dusty rings based on dust impact signals detected by Cassini RPWS. Icarus 279, 51–61.

Ye, S.–Y., Kurth, W. S., Hospodarsky, G. B., Averkamp, T. F. and Gurnett, D. A., 2016b. Dust detection in space using the monopole and dipole electric field antennas. J. Geophys. Res.: Space Physics, 121 (12), 11964–11972.

Ye, S.–Y., Kurth, W. S., Hospodarsky, G. B., Persoon, A. M., Sulaiman, A. H., Gurnett, D. A., Morooka, M., Wahlund, J.–E., Hsu, H.–W., Sternovsky, Z., Wang, X., Horanyi, M., Seiss, M., Srama, R., 2018. Dust observations by the Radio and Plasma Wave Science instrument during Cassini's Grand Finale. Geophys. Res. Lett. 45 (19), 101–110.

Ye, S.-Y., Vaverka, J., Nouzak, L., Sternovsky, Z., Zaslavsky, A., Pavlu, J., et al. (2019). Understanding Cassini RPWS antenna signals triggered by dust impacts. Geophysical Research Letters, 46, 10941– 10950. https://doi.org/10.1029/2019GL084150

Ye, S. Y., Averkamp, T. F., Kurth, W. S., Brennan, M., Bolton, S., Connerney, J. E. P., & Joergensen, J. L. (2020). Juno Waves detection of dust impacts near Jupiter. Journal of Geophysical Research: Planets, 125(6), e2019JE006367.

Zaslavsky, A., Meyer-Vernet, N., Mann, I., Czechowski, A., Issautier, K., Chat, G. L., Pantellini, F., Goetz, K., Maksimovic, M., Bale, S. D., & Kasper, J. C. (2012). Interplanetary dust detection by radio antennas: Mass calibration and fluxes measured by STEREO/WAVES. Journal of Geophysical Research: Space Physics (1978–2012), 117(A5), n/a-n/a. https://doi.org/10.1029/2011ja017480

Zaslavsky, A. (2015), Floating potential perturbations due to micrometeoroid impacts: Theory and application to S/WAVES data. J. Geophys. Res. Space Physics, 120: 855–867. doi: 10.1002/2014JA020635.

Zaslavsky, A., Mann, I., Soucek, J., Czechowski, A., Píša, D., Vaverka, J., ... & Vaivads, A. (2021). First dust measurements with the Solar Orbiter Radio and Plasma Wave instrument. Astronomy & Astrophysics, 656, A30.